# Proposed Cellular Network for Indian Conditions for Enhancement of Spectral Density and Reduction of Power Consumption & RF Pollution

Sumit Katiyar, Prof. R. K. Jain
Hi-Tech Institute of Engg. & Tech., Ghaziabad
Research Scholar, Singhania University,
Jhunjhunu, Rajasthan, India
sumitkatiyar@gmail.com, rkjain_iti@rediffmail.com

Prof. N. K. Agarwal Sr M IEEE
Inderprastha Engineering College, Sahibabad
Ghaziabad, India
agrawalnawal@gmail.com

*Abstract*— With the exponentially increasing demand for wireless communications the capacity of current cellular systems will soon become incapable of handling the growing traffic. Since radio frequencies are diminishing natural resources, there seems to be a fundamental barrier to further capacity increase. The solution can be found by using smart antenna systems.

Smart or adaptive antenna arrays consist of an array of antenna elements with signal processing capability that optimizes the radiation and reception of a desired signal, dynamically. Smart antenna can place nulls in the direction of interferers via adapting adaptive updating of weights linked to each antenna element. They thus cancel out most of the co-channel interference resulting in better quality of reception and lower dropped calls. Smart antenna can also track the user within a cell via direction of arrival algorithms. This paper focuses on about the smart antenna in hierarchical cell clustering (overlay-underlay) with demand based frequency allocation techniques in cellular mobile radio networks in INDIA.

I. INTRODUCTION

The demand of cellular radio service is growing rapidly, and in heavily populated areas the need arises to shrink the cell sizes and "scale" the clustering pattern. The extension of services into in-buildings and in pedestrian areas further enhances this trend. Cellular systems incorporate micro- and picocells for pedestrian use, with macro cells for roaming mobiles. This overall cellular environment needs a flexible network depending on particular situation / area. SDMA with overlay-underlay cell clustering (hierarchical cell clustering) is generally used for smoothing coverage in areas where topography or obstructions create shadowed regions, for accommodating regions with higher user concentrations, or for local private networks within the coverage of the public network.

The CDMA system features power control in each cell and a soft handoff, made possible by the usage of a single frequency throughout the system. The boundary between the cells is dynamic and obeys the soft handoff condition. Its location depends on the momentary threshold setting in each cell and on the transmission loss curves. Proper control of these parameters allows for the clustering of heterogeneous cells, sizing and shaping the cells, embedding a microcell in an umbrella cell, and also controlling the soft handoff zone. The threshold setting is controlled by desensitization, and is balanced by limiting the mobile transmitting power. The transmission power is controlled by smart antenna / intelligent antenna placing and beam shaping.

II. BASIC PRINCIPLES

A. *Smart Antenna / Adaptive Antenna*

Smart antenna is an array of antenna elements connected to a digital signal processor. Such a configuration dramatically enhances the capacity of a wireless link through a combination of diversity gain, array gain, and interference suppression. Increased capacity translates to higher data rates for a given number of users or more users for a given data rate per user. Multipath of propagation are created by reflection, diffraction and scattering. Also, interference signals such as that produced by the microwave oven are superimposed on the desired signals. Measurements suggest that each path is really a bundle or cluster of paths, resulting from surface roughness or irregularities. The random gain of the bundle is called multipath fading [1], [2], [3], [4], [5], [6], [7]. With the exponentially increasing demand for wireless communications the capacity of current cellular system will soon become incapable of handling the growing traffic. Since radio frequencies are diminishing natural resources, there seems to be a fundamental barrier to further capacity increase even in hierarchical cellular infrastructure. The solution can be found in smart antenna / adaptive antenna systems which utilizes SDMA approach. The principle of SDMA is quite different from the beam forming approaches. In fact operation of SDMA is analogous to that of human hearing (Intelligent System approach). As you can identify the direction from which the desired signal is coming with remarkable accuracy with the help of computer system inbuilt in adaptive antenna system. Because SDMA employs spatially selective transmission, an SDMA base station radiates much less total power than a conventional base station. One result is a reduction in network-wide RF pollution. Another is a reduction in power amplifier size. First, the power is divided among the elements, and then the power to each element is reduced because the energy is being delivered directionally. With a ten-element array, the amplifiers at each element need only transmit one-hundredth the power that would be transmitted from the corresponding single-antenna system [4].

Space Division Multiple Access (SDMA) can be used to multiply the capacity given by conventional multiple access techniques such as FDMA, TDMA and CDMA. However, the actual capacity gain which can be achieved with SDMA is highly dependent on the SDMA channel assignment and on the considered scenario (propagation, user distribution, traffic, mobility etc.). The vision of the "third generation" cellular systems incorporates micro and pico cells for pedestrians use and in buildings use, with macro cells for roaming mobiles. Connectivity between all these cells, while maximizing the total system capacity is the main challenge for engineers. Therefore channel assignment strategies for SDMA systems are to be analyzed. The available RF resources can be divided between macro, micro and pico cells. Therefore several channel assignment strategies for SDMA systems were discussed and some simulation results on the performance of SDMA systems were presented. The presented simulation results emphasize that inhomogeneous traffic causes a degradation of the SDMA capacity gain while dynamic inter-cellular channel allocation combined with SDMA can further increase the capacity as compared to pure SDMA and thus can enhance the efficiency of frequency usage.

*B. Principle of Working*

The smart antenna works as follows. Each antenna element "sees" each propagation path differently, enabling the collection of elements to distinguish individual paths within a certain resolution. As a consequence, smart antenna transmitters can encode independent streams of data onto different paths or linear combinations of paths, thereby increasing the data rate, or they can encode data redundantly onto paths that fade independently to protect the receiver from catastrophic signal fades, thereby providing diversity gain. A smart antenna receiver can decode the data from a smart antenna transmitter. This is the highest-performing configuration or it can simply provide array gain or diversity gain to the desired signals transmitted from conventional transmitters and suppresses the interference. No manual placement of antennas is required. The smart antenna electronically adapts to the environment by looking for pilot tones or beacons or by recovering certain characteristics (such as a known alphabet or constant envelope) that the transmitted signal is known to have. The smart antenna can also separate the signals from multiple users who are separated in space (i.e. by distance) but who use the same radio channel (i.e. center frequency, time-slot, and/or code); this application is called Space-division multiple access (SDMA) [17].

Beam forming presents several advantages to antenna design .Firstly, space division multiple access (SDMA) is achieved since a beam former can steer its look direction towards a certain signal. Other signals from different directions can reuse the same carrier frequency.

Secondly, because the beam former is focused in a particular direction, the antenna sensitivity can be increased for a better signal to noise ratio, especially when receiving weak signals. Thirdly, signal interference is reduced due to the rejection of undesired signals.

One of the foremost advantages offered by the software radio technology is flexibility. Because beam forming is implemented in software, it is possible to investigate a wide range of beam forming algorithms without the need to modify the system hardware for every algorithm.

*C. Merits of Smart Antenna*

Smart antennas at base stations can be used to enhance mobile communication systems in several ways:
- Increased BS range
- Less interference within the cell
- Less interference in neighboring cells
- Increased capacity by means of SFIR or SDMA

The influence of traffic scenarios and channel allocation schemes on the capacity performance of SDMA systems has been addressed in [8]. The presented simulation results emphasize that inhomogeneous traffic causes a degradation of the SDMA capacity gain while dynamic inter-cellular channel allocation combined with SDMA can further increase the capacity as compared to pure SDMA and thus can enhance the efficiency of frequency usage. 'Smart' antenna transmitters emit less interference by only sending RF power in the desired directions. Furthermore, 'smart' antenna receivers can reject interference by looking only in the direction of the desired source. Consequently 'smart' antennas are capable of decreasing CCI. A significantly reduced CCI can be taken as advantage of *Spatial Division Multiple Access* (SDMA) [9]. The same frequency band can be re-used in more cells, i.e. the so called frequency re-use distance can be decreased. This technique is called *Channel Re-use via Spatial Separation*.

*D. Spatial Division Multiple Access (SDMA)*

Adaptive antenna also allows base station to communicate with two or more mobiles on the same frequency using SDMA. In SDMA multiple mobiles can communicate with single base station on same frequency by using highly directional beams and/or forming nulls in the direction of all but one of the mobile on a frequency. The base station creates multiple channels using the same frequency but separate in space [9].

*E. Applications in Mobile Communications*

A space-time processor ('smart 'antenna') is capable of forming transmit/receive beams towards the mobile of interest. At the same time it is possible to place spatial nulls in the direction of unwanted interferences. This capability can be used to improve the performance of a mobile communication system [17].

III. HIERARCHICAL ARCHITECTURE

Splitting cells into smaller cells can reduce the frequency reuse distance to improve network capacity with in a certain area other than increasing the cost of the fixed infrastructure; cell splitting also causes the problem of increasing handoff rate and even the handoff failure rate when high speed users roam in the network. To solve this problem larger cells are overlaid on these smaller cells and different classes of users (usually classified by speed) are initially assigned to the proper types of cells (that in proper tier). We call this kind of cellular network a hierarchical cellular network [10]. Macro cells are used for high speed users, micro cells for low speed users (pedestrians) and pico cells for in-building use or stationary users. The radio resources are distributed among these cells based on candidate cell location with corresponding costs, the amount of available bandwidth, the maximum demand for service in each geographical area and the revenue potential in each customer area. Based on these data, the model determines the size and location of the cells and the specific channel to be allocated to each cell.

Hierarchical wireless overlay network allow for flexible mobility management strategies and quality of service differentiation. One of the crucial problems in hierarchical overlay network is the assignment of wireless data uses to the different layers of overlay architecture [11]. Another approach is the "velocity-sensitive" cell selection method. In this approach, the cell assignment decision is made based on the mobile's estimated velocity, so that slow mobiles can be assigned to microcells, while fast mobiles are assigned to macrocells. In [12], all newly arrived users are assigned to a microcell by default. When a user moves out of the coverage of his / her current cell, the cell dwell time is compared with a predetermined threshold. If a user is slowly moving (i.e., if the cell dwell time is larger than the threshold), the user is handed over to a neighboring microcell, otherwise the user is handed over to the macrocell. In [13], the channel condition information is utilized in conjunction with mobile velocity information. Essentially, the cell selection is done by comparing the signal strength of the microcell and the macrocell, so that the cell with the stronger signal is selected. A certain negative offset is applied to the signal strength of the microcell, which becomes effective when the mobile enters the coverage of microcells and is reduced over time. Due to this negative offset, mobiles are more likely to select the macrocell at first, if the signal strengths of macrocells and microcells are comparable. However, as the offset decreases, more mobiles select the microcell if they remain in the coverage of microcells. As a result, slow mobiles would choose microcells, while the fast mobiles choose the macrocell.

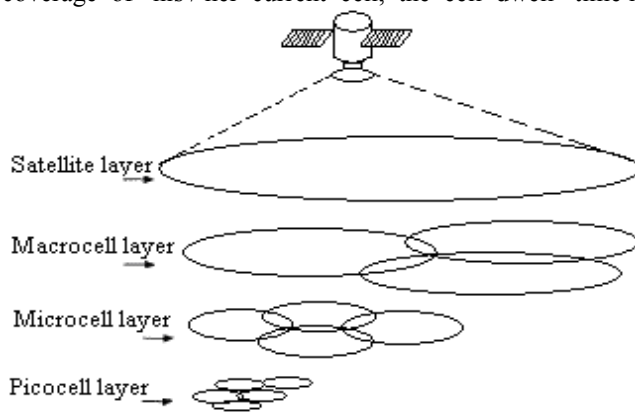

Figure 1.  Hierarchical Cellular Network

IV. PROPOSED NETWORK

It has been proved beyond doubt that spectral density can be enhanced manifold using hierarchical cell clustering (overlay-underlay) networks with SDMA and dynamic frequency allocation (DFA). All above requirements can be achieved through employment of smart antenna / intelligent antenna at base stations. Proper employment of smart antenna not only enhances the spectral density but also help in providing solutions for variable traffic requirements. The traffic density is a variable function. Equipping base stations with adaptive (smart) antenna arrays will enable beam steering in contrast to omni directional or sectorized single antenna systems. Thus it becomes possible for base stations to radiate directed to specific users on the downlink as well as receive directed on the up-link. This can be exploited by reusing channels within a cell for mobile stations (MS) which are spatially separable by the smart antenna array. As a result demand based RF resources allocation is possible and this provide flexibility in the cellular networks for better utilization of RF resources. With a use of smart antenna at base station this has also been proved beyond doubt that power consumption will be reduced with a use of low power transmitters. As a result energy resources will be saved. Since electricity is a scarce resource in INDIA and most of the time in rural / semi urban areas base stations are being run by DG sets. In metros also DG sets are being used as backup load. Cellular base station diesel consumption is around 10% of the total diesel consumption in INDIA which can be reduced to 1% through proper networking and employment of smart antenna at base stations. In addition to this, RF pollution has been considered as health hazard for man kind. With the use of smart antenna at base stations the RF pollution can be reduced manifold; this has also been proved beyond doubt [15]. Taking into consideration all above advantages of smart antenna and prevailing circumstances in INDIA, it is strongly recommended that new 3G systems and extension of 2G and 2.5G systems should be implemented with proper planning of networking. All advantages of proved technologies should be utilized while finalizing the networking.

- Deployment along city streets
- Coverage along winding roads
- Coverage under the ground (subway coverage / tunnel / basement / underground metro railway systems etc).
- In-building communication systems as per requirement (fig 2).
- Micro- and Picocell deployment along with DFA and SDMA approach (fig 3).
- Base station may be equipped with smart antenna in hierarchical cell clustering (overlay-underlay).

We suggest in this paper, a practical cellular system (fig 4) which will enhance spectral efficiency and reduce RF pollution & power consumption of existing 2 G & 2.5 G system with small change in the existing 2G & 2.5G systems. Existing sectored antenna will be replaced through smart antenna with SDMA approach.

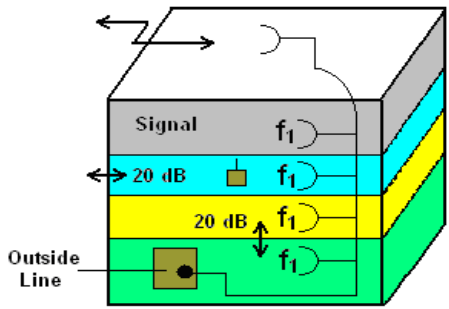

Figure 2. Signal within a building

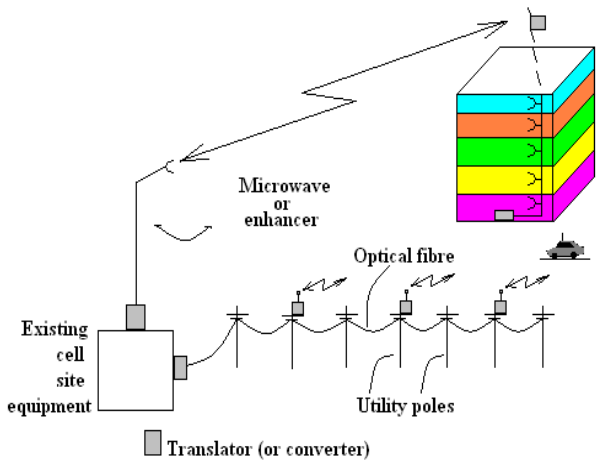

Figure 3. Microcell delivery system

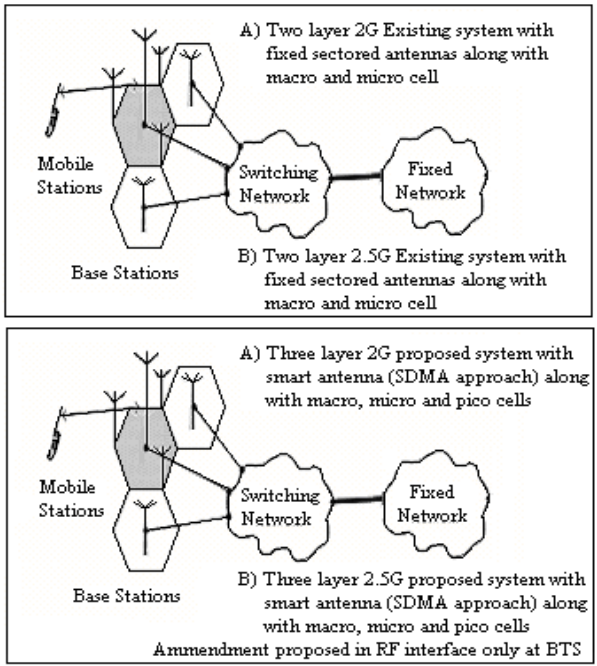

Figure 4. Proposed Network

Mobile telephone density is increasing in metro cities out of proportion. The installed network is not capable to meet the requirements. We can take an example of Indirapuram, locality of Ghaziabad where high rise buildings are coming up very fast and spectral density requirement is very high for a small area. The solution is to add new spectrum (frequency bands) to meet the requirements or to find out ways and means to use the same frequency again and again by reducing interference through low power transmitters and proper networking. Spectral density can be enhanced by use of smart antenna along with DFA. (Demand based Frequency Allocation). Due to lot of hindrance between transmitters and receivers the losses and interference due to multipath are very high. For meeting this requirement hierarchical system is the only solution with low power transmitters so that in-building subscribers, pedestrian subscribers and low speed traffic may be covered by microcell. Multiple requirements (such as fast traffic, slow traffic, pedestrians, In-building subscribers, malls subscribers and market subscribers etc.) all are to be connected with the same system i.e. hierarchical cellular system. Wherever density requirement (malls, market, railway station etc.) is high, the solution is picocell which will provide additional resources in absence of additional spectrum. However microcell can cover slow traffic, pedestrians and In-building subscribers (with the help of In-building solution). It will be appropriate to mention that vendors in INDIA are not using available technology for meeting the requirements. Instead of this, they are adding spectrum and base stations with high power transmitters for meeting the requirements. Very few vendors are using In-building solution. Instead of this, they are using high power microcells for penetrating signal inside the building which is consuming more power and increasing level of pollution. In this paper, we have suggested simple technologies with proper planning of networking for meeting multiple requirements. Figure 2 and figure 3 illustrate In-building solution & Microcell delivery system respectively [14].

In order to increase the mobile communication system capacities without jeopardizing the system quality of service, smart antenna with DFA is suggested. A paper on "WCDMA Forward Link Capacity Improvement by Using Adaptive Antenna with Genetic Algorithm Assisted MDPC (Minimum Downlink Power Consumption) Beamforming Technique has been published" [16]. The simulated result clearly shows that adaptive antenna can be used for enhancing the mobile communication system capacity.

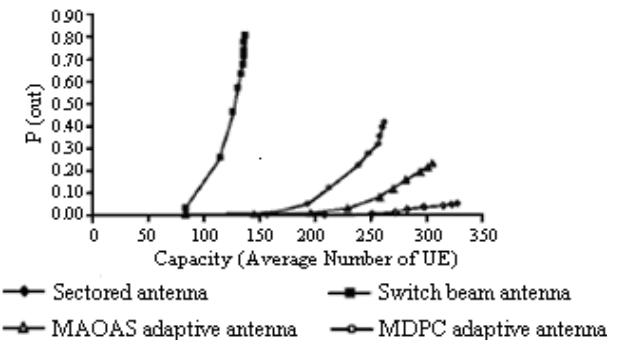

Figure 5. Average number of UE under different outage probability with different types of antenna systems

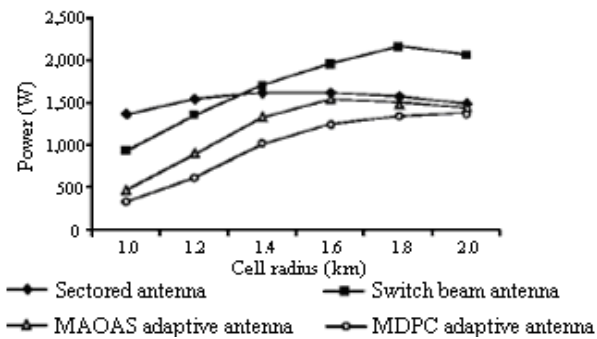

Figure 6. Outage probability by using different types of antenna systems at different cell size

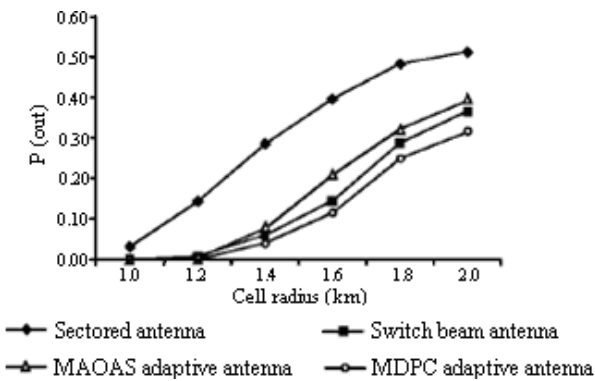

Figure 7. Outage probability by using different types of antenna systems at different cell size

## V. DISCUSSION

Smart Antennas are currently suggested in selected base station with coverage of interference problems. Range extension is particularly important in PCS systems at 1.9 GHz because the propagation loss is higher at 900 MHz. In future, for capacity increase in TDMA systems, adaptive arrays with four antenna elements in IS-136 or multibeam antennas with four or eight beam in GSM may be deployed at all base stations within a cellular system region to decrease the frequency reuse factor from 7 to 4, nearly doubling capacity. However a practical model can be developed taking an example of Indirapuram, Ghaziabad, where multiple requirements suggest too many multi stories buildings, malls, highways traffics (fast traffic), In-building subscribers, pedestrians, slow traffic etc. are to be addressed. After proper survey of signal strength and traffic density at a different point's optimal network model can be developed applying macro, micro, pico cells in hierarchical structures using smart antenna and SDMA approach clubbed with DFA. All advantages of proved easily implemented technologies stated above may be utilized while finalizing optimal network for addressing multiple requirements. For mobile communications, the directivity of the antennas has to be adaptive such that the beams pointing to the users can follow their movements. This new technique adaptive SDMA can be utilized. In essence, the scheme can adapt the frequency allocations to where the most users are located.

## VI. CONCLUSION

In this communication the influence of traffic scenarios in INDIA and demand based channel allocation schemes in hierarchical cellular networks along with performance of SDMA systems using smart antenna has been addressed. Based on analysis of proved technologies, a simple cellular networking approach for Indian conditions has been recommended. It is clear beyond doubt that proposed network will enhance spectral efficiency and reduce power consumption as well as RF pollution.